\documentclass[12pt]{iopart}

\usepackage{epsfig,mathrsfs,graphicx,amssymb}
\def\rmD{\mathrm D}

\begin{document}

%%% remove comment delimiter ('%') and select language if required
%\selectlanguage{spanish} \begin{document}
\title[Nonclassical time correlation functions]{Nonclassical time correlation functions in continuous quantum measurement}

\author{Adam Bednorz$^1$, Wolfgang Belzig$^2$ and Abraham Nitzan$^3$}
%\altaffiliation[Electronic address: ]{Adam.Bednorz@fuw.edu.pl}
\address{$^1$ Faculty of Physics, University of Warsaw, Ho\.za 69, PL-00681 Warsaw, Poland}

\address{$^2$ Fachbereich Physik, Universit{\"a}t Konstanz, D-78457 Konstanz, Germany}

\address{$^3$ Raymond and Beverly Sackler Faculty of Exact Sciences, School of Chemistry, Tel-Aviv University, Tel-Aviv 69978, Israel}
\ead{Adam.Bednorz@fuw.edu.pl}

\begin{abstract} 
  A continuous projective measurement of a quantum system often
leads to a suppression of the dynamics, known as the Zeno effect. Alternatively,
generalized nonprojective, so-called ``weak'' measurements can be carried
out. Such a measurement is parameterized by its strength parameter that
can interpolate continuously between the ideal strong measurement with no
dynamics—the strict Zeno effect, and a weak measurement characterized by
almost free dynamics but blurry observations. Here we analyze the stochastic
properties of this uncertainty component in the resulting observation trajectory.
The observation uncertainty results from intrinsic quantum uncertainty, the effect
of measurement on the system (backaction) and detector noise. It is convenient to
separate the latter, system-independent contribution from the system-dependent
uncertainty, and this paper shows how to accomplish this separation. The system-dependent
uncertainty is found in terms of a quasi-probability, which, despite its
weaker properties, is shown to satisfy a weak positivity condition. We discuss
the basic properties of this quasi-probability with special emphasis on its time
correlation functions as well as their relationship to the full correlation functions
along the observation trajectory, and illustrate our general results with simple
examples.We demonstrate a violation of classical macrorealism using the fourth-order
time correlation functions with respect to the quasi-probability in the twolevel
system.
\end{abstract}

\submitto{\NJP}
\maketitle

\section{Introduction}

The continuous projective von Neumann quantum measurement
\cite{neumann} leads to a suppression of the dynamics, known as the
quantum Zeno effect (QZE) \cite{zeno}. To escape this problem, modern
quantum measurement theory offers generalizations of the projective
measurement to so-called positive operator-valued measures (POVM)
\cite{povm,wimi}, where a projection is replaced by a softer operation such
as a Kraus operator \cite{kraus}.  Such operators can describe not
only projective measurements but also weak measurement, in which case
the action of the POVM leaves the state almost unchanged. By virtue of
the Naimark theorem \cite{naim} POVMs are equivalent to projective
measurements in an extended Hilbert space that includes additional
detector degrees of freedom. The effect of a continuous application of
Kraus operators, which correspond to a time-continuous measurement can
be described by stochastic evolution equations such as Lindblad-type
equations \cite{lind} for the system density matrix or Langevin
equations for individual system trajectories, physically describing
irreversible effects like decoherence and decay affected by the
measurement process.

Weak measurements\cite{aav} make it possible to escape the QZE by
paying a price in terms of an imperfect detection. In the extreme case
the dynamics of the system is (almost) free but the measurement
outcome is obscured by a large detection noise. This is similar to the
problem of a quantum linear amplifier, which can amplify both
complementary noncommuting observables, like $\hat{x}$ and $\hat{p}$,
but only if accompanied by a large noise\cite{caves}. The
interpretation of weak measurements of correlation functions is
sometimes paradoxical: one must either accept unusually large values
of the physical quantity\cite{aav} or replace the probability by a
quasiprobability \cite{quasi}. Weak measurements are also very useful
in quantum feedback protocols \cite{kor}.

The QZE lies at the strong limit of a spectrum of measurements whose
strong/weak character can be changed continuously \cite{qnd, cav1,
  gag, aud, men,kor2,hof,jast}, using e.g. a Gaussian POVM
\cite{gpovm,cav2}. The Gaussian POVM is also the key element of the
continuous collapse interpretation of quantum mechanics \cite{coll}.
These models lead to various types of expressions for time correlation functions \cite{zoll}.
Here, for the first time we use Gaussian POVM for continuous
measurement to describe higher order symmetrized time correlation functions.  Such functions
are known for the two-time case of a harmonic system \cite{gpovm} or in the weak measurement limit 
for the two-level system \cite{paz}
(but not in general) and are necessary to explain the recent
experiment that shows nonclassical behavior of time correlation functions in a two-level system \cite{este}.  
The calculations are faciliated by making a
deconvolution of the outcome time trace probability into the
probability component associated with the white detection noise and a
quasiprobability that describes the intrinsic system uncertainty. Such a
deconvolution has the advantage that we can make use of basic
properties of the quasiprobability, e.g., the weak positivity
\cite{wpp}, which states that the second order correlation function
matrix is positive definite. Our scheme provides a unified and concise
treatment of weak measurements and the QZE, pointing out the general
trade-off between measurement and decoherence.  By comparing the
average signal to the associated noise we also establish limits on the
uncertainty of the outcome and its dependence on measurement
characteristics.

The time correlation functions obtained by our approach provide a convenient route for the analysis of uncertainty properties of systems undergoing weak measurements. 
Taking a two-level system as an example, a single nondemolishing
measurement of an observable not commuting with the Hamiltonian is not
possible in both time and frequency domain although the latter gives a
better signal-to-noise ratio. Although this is intuitively clear,
using our approach, it is possible to establish and compare bounds on
the outcome uncertainty. For another simple example, the continuous
position measurement of a harmonic oscillator, we show that the same
measurement procedure does not lead to the QZE. Instead, the
continuous measurement leads to unbounded growth in noise, in
agreement with the general observation \cite{cav2} and in analogy to
anti-Zeno effect\cite{aze}.

The proposed separation has another important consequence. If we assume classical macrorealism in quantum
mechanics then the statistics of the outcomes with the detection noise
subtracted in the limit of noninvasive measurement should correspond
to a positive definite probability.  In contrast, we show that  macrorealism assumption is
violated by demonstrating that our quasiprobability is somewhere
negative. Such violation has been recently demonstrated experimentally \cite{este}.
In fact, if we additionally assume dichotomy or boundedness of
the quantum outcomes, the violation can occur already on the level of
second order correlations of a single observable as shown by Leggett, Garg and others
\cite{paz,leg}, but also indirectly, subtracting the unavoidable (and necessarily divergent) noise \cite{kobr}.  
However, as follows from weak positivity,
without these additional assumptions, second order correlations are
not sufficient to violate macrorealism. Instead, one needs at least
fourth order averages to see this violation.  In this paper we demonstrate, that a special
fourth order correlation function in the two-level system, reminiscent
of the Leggett-Garg proposal\cite{leg}, can reveal the negativity of
the quasiprobability in this case and consequently can be used to
violate macrorealism, without having to make any additional assumptions.

The paper is organized as follows. We first define the continuous
Gaussian POVM and obtain the probability distribution for the
continuously measured observable. We then make the deconvolution of
this probability into detection noise and a quasiprobability and
introduce a formalism for evaluating time correlation functions. With
that we are able to prove the weak positivity. We show that the time
evolution associated with the quasiprobability can be formulated
either as a quantum Langevin equation driven by a white Gaussian
noise, or a Lindblad-type master equation for the non-selective system
density matrix. We also show how the required time correlation
functions can be calculated from these stochastic equations. Next, we
demonstrate the general trade-off between dynamics and measurement,
taking a two-level system as example, and
discuss the behavior of the average signal and the noise in these
prototype system.  Then we construct the Leggett-Garg inequality
without assuming dichotomy or boundedness of the measurement
variable. 
Finally, for completeness, we discuss the harmonic oscillator case and show how
and when the Zeno effect emerges within our
formalism. Several instructive proofs of formulas are presented in
Appendices.

\section{Quasiprobability and weak positivity}

We begin by introducing a general scheme of continuous measurement and
describe its properties. For a given system characterized by a
Hamiltonian $\hat{H}$ and an initial system state given by a density
matrix $\hat{\rho} $, we consider the measurement of one, generally
time-dependent, observable $\hat{A}$. A description amenable to
continuous interpolation between hard and soft measurement can be
formulated in terms of Kraus operators \cite{kraus, aav}. We assume a
Gaussian form of the Kraus operators, whereupon the state of the
system following a single instantaneous measurement is given by
\begin{equation} \label{rho01} %\label{ZEqnNum505131} 
\hat{\rho }_{1} (a)=\hat{K}(a)\hat{\rho } \hat{K}(a),
\end{equation} 
\begin{equation} \label{kra1} 
\hat{K}(a)=(2\bar{\lambda }/\pi )^{1/4} \rme^{-\bar{\lambda }(a-\hat{A})^{2} } . 
\end{equation} 
Note that in (\ref{rho01}) the non-negative definite operators
$\hat{\rho } $ and $\hat{\rho }_{1} $ represent the states of the
system just before and just after the measurement. The probability
that the measurement of $\hat{A}$ gives the outcome $a$ is given by
\cite{povm}
\begin{equation} \label{prob} 
P(a)={\rm Tr}\hat{\rho }_1(a),
\end{equation} 
which is normalized, $\int da\; P(a)=1$. The Kraus operator
(\ref{kra1}) depends on the parameter $\bar{\lambda }$, which
characterizes the weakness of the measurement. For $\bar{\lambda }\to
\infty $, we recover a strong, projective measurement with an exact
result but a complete destruction of coherence, while $\bar{\lambda
}\to 0$ corresponds to a weak measurement with almost no influence on
the state of the system, $\hat{\rho }_1(a) \sim \hat{\rho }$, but very
large measurement uncertainty of the order of $\sim 1/\bar{\lambda }$.
The probability distribution (\ref{prob}) is consistent with
projective measurement scheme, namely $\langle
a\rangle=\mathrm{Tr}\hat{A}\hat{\rho}$.

Let us imagine a continuous
sequence of meters interacts with the system. The meters are prepared
with a Gaussian wave function, the interaction is proportional to the
product of the system observable $\hat{A}$ and the meter momentum, and the
position of each meter is read out after the interaction 
The post-interaction position of the meters is the
measurement result $a(t)$\cite{wimi,jast,zoll,gpovm}.

Repeated measurements of this type can be described by applying such
Kraus operators sequentially, separated by time steps $\Delta t$.  In
what follows we make the reasonable assumption that for a given
measuring device (``meter'') the weakness parameter $\bar{\lambda }$
is inversely proportional to the measurements frequency, i.e.,
\begin{equation} \label{4)} 
\bar{\lambda }=\lambda \Delta t 
\end{equation} 
with constant \textit{$\lambda $}. In the continuum limit,
$\bar{\lambda },\Delta t\to 0$, we obtain (Appendix A) the Kraus
operator as a functional of $a(t)$
\begin{equation} \label{kra2} 
  \hat{K}_{h} [a(t)]\equiv \rme^{\left(\rmi/\hbar \right)\hat{H}t}
  \hat{K}\left[a\left(t\right)\right]=C{\rm {\mathcal T}}\rme^{-\int
    \lambda (a(t)-\hat{A}(t))^{2} \rmd t}   
\end{equation} 
where $a(t)$ is the measurement outcome, $\hat{A}(t)$ is the operator
$\hat A$ in the Heisenberg representation with respect to the
Hamiltonian $\hat{H}$, $\hat{A}\left(t\right)=\exp
\left(\rmi\hat{H}t/\hbar\right)\hat{A}\exp \left(-\rmi
  \hat{H}t/\hbar\right)$, ${\rm {\mathcal T}}$denotes time ordering
(later times on the left) and $C$ is a normalization factor. Note that
$\hat{K}_{h}[a] $ is the Heisenberg representation of
$\hat{K}[a]$. The analog of (\ref{prob}) is the functional probability
\begin{equation} \label{pkk} 
P[a]={\rm Tr}\hat{K}^{\dag } [a]\hat{K}[a]\hat{\rho }  
\end{equation} 
which satisfies the normalization $\int \rmD a{\kern 1pt}
P[a]=1$. Whenever some functional measure $\rmD$ is introduced here, we
tacitly include all proper normalization factors in it.

It is convenient to write (\ref{kra2}) as a Fourier transform 
\begin{equation} \label{7)} 
  \hat{K}_{h} [a]=\int  \rmD\phi {\rm {\mathcal T}}\rme^{\int \rmd t[\rmi\phi
    (t)(\hat{A}(t)-a(t))-\phi ^{2} (t)/4\lambda ] }  
\end{equation} 
so that
\begin{eqnarray} \label{8)} 
  && P[a]={\rm Tr}\left(\hat{K}\left[a\right]\hat{\rho }
    \hat{K}\left[a\right]\right)
 ={\rm Tr}\left(\int  \rmD\phi _{+} {\rm {\mathcal T}}\rme^{\int \rmd t[\rmi\phi
     _{+} (t)(\hat{A}(t)-a(t))-\phi _{+} ^{2} (t)/4\lambda ] }\times\right.
     \nonumber\\
     &&\left.
   \hat{\rho } 
   \int  \rmD\phi _{-} \tilde{{\rm {\mathcal T}}}\rme^{\int \rmd t[\rmi\phi
     _{-} (t)(\hat{A}(t)-a(t))-\phi _{-} ^{2} (t)/4\lambda ] } \right) 
\end{eqnarray} 
where $\tilde{{\mathcal T}}$denotes inverse time ordering (later times
on the right). Changing integration variables according to $\chi
=\phi _{+} +\phi _{-} $ and $\phi =(\phi _{+} -\phi _{-})/2$ we can
write
\begin{eqnarray} \label{prob2} 
&&P[a]=\int  \rmD\phi {\kern 1pt} \rme^{-\int \rmd t\phi ^{2} (t)/2\lambda  } 
\int \rmD\chi {\kern 1pt} \rme^{-\int \rmd t\chi ^{2} (t)/8\lambda  } \rme^{-\int \rmi\chi (t)a(t)\rmd t }\\
&& \times {\rm Tr}\; {\mathcal T}\rme^{\int \rmi(\chi (t)/2+\phi
  (t))\hat{A}(t)\rmd t } \; \hat{\rho }\; \tilde{{\rm {\mathcal
      T}}}\rme^{\int \rmi(\chi (t)/2-\phi (t))\hat{A}(t)\rmd t } 
\nonumber
\end{eqnarray} 
The last line can be written alternatively as (see Appendix B) 
\begin{equation} \label{hphi} 
  {\rm Tr}\; {\mathcal T}\rme^{\rmi\int \chi (t)\hat{A}_{\phi }
    (t)\rmd t/2 } \; \hat{\rho }\; \tilde{\mathcal T}\rme^{\rmi\int \chi
    (t)\hat{A}_{\phi } (t)\rmd t/2 }   
\end{equation} 
where $\hat{A}_{\phi } (t)$ denotes the operator$\hat{A}$ in a
modified Heisenberg picture, namely with respect to the Hamiltonian
$\hat{H}-\hbar \phi (t)\hat{A}$.

(\ref{prob2}) and (\ref{hphi}) describe the outcome of the
continuous measuring process in terms of the probability distribution
functional $P\left[a\left(t\right)\right]$ of the observation function
$a(t)$.  This distribution reflects the quantum uncertainty, the
modified system time evolution caused by the measurement (the
backaction effect) and the uncertainty associated with the weak
measurement that can be thought of as reflecting detector noise. A
more transparent view of these contributions is obtained by separating
the latter, system independent contribution from the system dependent
effects. This is achieved by considering the moment generating
functional (MGF) $M[\chi]=\rme^{S\left[\chi \right]} $, where
$S\left[\chi \right]$ is the cumulant generating functional (CGF),
given by
\begin{eqnarray} \label{cgf} 
&&M[\chi]=\rme^{S[\chi ]} =\int  \rmD a\;\rme^{\rmi\int \chi (t)a(t) } P[a]=\\
&&\rme^{-\int \rmd t\chi ^{2} (t)/8\lambda  } \int  \rmD\phi {\kern 1pt} 
\rme^{-\int \rmd t\phi ^{2} (t)/2\lambda  }
{\rm Tr}\; {\mathcal T}\rme^{\rmi\int \chi (t)\hat{A}_{\phi } (t)\rmd t/2 } \; \hat{\rho }\; \tilde{{\mathcal T}}\rme^{\rmi\int \chi (t)\hat{A}_{\phi } (t)\rmd t/2 }. \nonumber
\end{eqnarray} 
The CGF can be divided into two parts $S[\chi ]=S_{d}
[\chi ]+S_{q} [\chi ]$ with
\begin{equation} \label{sdd} 
S_{d} [\chi ]=-\int  \rmd t\chi ^{2} (t)/8\lambda  
\end{equation} 
and
\begin{equation} \label{sqq} 
\rme^{S_{q} [\chi ]} =\int  \rmD\phi {\kern 1pt} e^{-\int \rmd t\phi ^{2} (t)/2\lambda  }
{\rm Tr}\; {\mathcal T}\rme^{\rmi\int \chi (t)\hat{A}_{\phi } (t)\rmd t/2 } \; \hat{\rho }\; \tilde{{\mathcal T}}\rme^{\rmi\int \chi (t)\hat{A}_{\phi } (t)\rmd t/2 }.\nonumber
\end{equation} 
Note that $S[0]=S_{d} \left[0\right]=S_{q} \left[0\right]=0$. On the
level of probabilities this decomposition corresponds to the
convolution
\begin{equation} \label{conv} 
P[a]=\int  \rmD a'P_{d} [a-a']P_{q} [a'] ,
\end{equation} 
where 
\begin{equation} \label{pdd} 
P_{d} \left[a\right]=\int  \rmD \chi \rme^{\int  \rmd t(\chi (t)a(t)/\rmi- \chi ^{2} (t)/8\lambda) }\propto \rme^{-2\lambda \int a^{2}(t)\rmd t }  
\end{equation} 
corresponds to a Gaussian noise with zero average and correlation
$\langle a(t)a(t')\rangle _{d} =\delta (t-t')/4\lambda $ that may be
interpreted as the noise associated with the detector, and where
\begin{equation} \label{pqq} 
P_{q} [a]=\int  \rmD\chi {\kern 1pt} e^{-\int  \rmi\chi (t)a(t)\rmd t} 
\rme^{S_{q} [\chi ]}  
\end{equation} 
is a distribution associated with the intrinsic system uncertainty as
well as the measurement backaction. It is normalized, $\int \rmD a\, P_{q}
[a]=1$, but not necessarily positive, and will be referred to as a
quasi-probability \cite{quasi,schleich,tsang}.  In the limit of weak,
noninvasive measurement, $\lambda\to 0$, $P_d$ diverges while $P_q$
has a well defined limit
\begin{equation} \label{sqq0} 
\rme^{S_{q} [\chi ]} \stackrel{\lambda \to 0}{\longrightarrow}
{\rm Tr}\; {\rm {\mathcal T}}\rme^{\rmi\int \chi (t)\hat{A} (t)\rmd t/2 } \; \hat{\rho }\; \tilde{{\mathcal T}}\rme^{\rmi\int \chi (t)\hat{A}(t)\rmd t/2 }.
\end{equation} 

Consider now this distribution $P_q$. First note that while it is not a real
probability functional, it is possible to calculate moments
$\langle\rangle$ and cumulants $\langle\langle\rangle\rangle$ with
respect to this measure as partial derivatives of the quasi-CGF, respectively
\begin{eqnarray} \label{avgs} 
&&\langle a(t_{1})\cdots a(t_{n})\rangle _{q} =\left. \frac{\delta ^{n} \exp S_{q} [\chi] }{\delta \rmi\chi \left(t_{1} \right)\cdots\delta \rmi\chi \left(t_{n} \right)} \right|_{\chi =0} , \\
&&\langle\langle a(t_{1})\cdots a(t_{n})\rangle\rangle _{q} =\left. \frac{\delta ^{n} S_{q}[\chi] }{\delta \rmi\chi \left(t_{1} \right)\cdots\delta \rmi\chi \left(t_{n} \right)} \right|_{\chi =0}\nonumber.
\end{eqnarray} 
In particular, for $t_n\ge \dots\ge t_2\ge t_1$,
\numparts
\begin{eqnarray}
 &&
\langle a(t)\rangle _{q} =\int  \rmD_{\lambda } \phi {\kern 1pt} {\rm Tr}[\hat{A}_{\phi } (t)\hat{\rho }],
\label{av1}\\ %      (\label{ZEqnNum589959}a)
&&\langle a(t_1)a(t_2)\rangle _{q} =
\int  \rmD_{\lambda } \phi {\kern 1pt} {\rm Tr}[\{ \hat{A}_{\phi } (t_1),\hat{A}_{\phi } (t_2)\} \hat{\rho }]/2,\label{av2}\\ %    (\eqref{ZEqnNum589959}b)
&&\langle a(t_1)a(t_2)a(t_3)\rangle _{q} =\int  \rmD_{\lambda } \phi {\kern 1pt} \label{av3}
{\rm Tr}[\{ \hat{A}_{\phi } (t_1),\{ \hat{A}_{\phi } (t_2),\hat{A}_{\phi } (t_3)\} \} \hat{\rho }]/4,
\\ %  (\eqref{ZEqnNum589959}c) 
&&\langle a(t_1)\cdots a(t_n)\rangle _{q} =\int  \rmD_{\lambda } \phi {\kern 1pt}
{\rm Tr}[\{ \hat{A}_{\phi } (t_1),\{
\{\hat{A}_{\phi } (t_2),\cdots\hat{A}_{\phi } (t_n)\}\cdots \} \hat{\rho }]/2^{n-1},
\nonumber
\end{eqnarray}
\endnumparts
\noindent (see Appendix C), where we have defined $\rmD_{\lambda } \phi
=\rmD\phi {\kern 1pt} \rme^{-\int \rmd t\; \phi ^{2} (t)/2\lambda } $. Here and
below we use standard notation $\{ \hat{A},\hat{B}\}
=\hat{A}\hat{B}+\hat{B}\hat{A}$ and
$[\hat{A},\hat{B}]=\hat{A}\hat{B}-\hat{B}\hat{A}$. 

Secondly, from Eq.~(\ref{av2})
follows the important so-called \emph{weak positivity} property of second order
correlations \cite{wpp}
\begin{equation} \label{wepo} 
\langle F^{2} [a]\rangle _{q} =\int  \rmD\phi {\kern 1pt} e^{-\int  \rmd t\phi ^{2} (t)/2\lambda } {\rm Tr}F^{2} [\hat{A}_{\phi } ]\hat{\rho }\ge 0 
\end{equation} 
for $F[a]=\int dt(f(t)a(t)+g(t))$ and arbitrary functions $f$ and $g$.
It can be interpreted as a generalization of the Robertson-Schr{\"o}dinger uncertainty principle \cite{robsch}.
This property has an important implication that no test based
solely on maximally second order correlations can reveal the
negativity of the quasiprobability.  First and second order
correlations can be represented by a completely classical, positive
Gaussian probability distribution
 \begin{equation}
 P'_q[a]\propto \exp\left(-\int \rmd t \rmd t' \delta a(t)f^{-1}(t,t')\delta a(t')/2\right),
 \end{equation}
 where $\delta a(t)=a(t)-\langle a(t)\rangle_q$, $f(t,t')=\langle
 \delta a(t)\delta a(t')\rangle_q$ and $f^{-1}$ is its inverse defined
 by $\int \rmd t f(t',t)f^{-1}(t,t'')=\delta(t'-t'')$.  The weak
 positivity guarantees that both $f$ and $f^{-1}$ are positive
 definite and consequently $P'_q$ is a correct real probability
 distribution.  To check that $P_q$ differs from $P'_q$ and
 demonstrate its negativity one needs higher order correlations or
 additional assumptions (e.g. boundedness or dichotomy of $a$ as it
 happens in Leggett-Garg inequality \cite{leg}).

 To end this Section, we consider the special case in which the
 Hamiltonian commutes with $\hat{A}$ (or the noncommuting part is
 negligible during the interesting timescale). Furthermore, let us
 take the initial state of the system to be an eigenstate $| a\rangle$
 of $\hat{A}$, i.e. $\hat{\rho }\left(t=0\right)=| a\rangle\langle a|
 $, $\hat{A}|a\rangle=a|a\rangle$. Consider a measurement performed
 during the time interval $t_{0} $,
\begin{equation} \label{ava} 
\bar{a}=(1/t_0)\int _{0}^{t_{0} } \rmd t\;a(t)
\end{equation} 
In this case we find (Appendix D) that $\langle \bar{a}\rangle
=\langle \bar{a}\rangle _{q} =a$ and $\langle (\delta \bar{a})^{2}
\rangle =\langle (\delta \bar{a})^{2} \rangle _{d} =1/4\lambda t_{0} $
with $\delta X=X-\langle X\rangle$. We can see the intuitively
expected effect of an increasing measurement duration to lead to an
improved signal to noise ratio with time, which goes as
\begin{equation} \label{21)} 
\frac{\langle \bar{a}\rangle }{\sqrt{\langle (\delta \bar{a})^{2} \rangle } } =2a\sqrt{\lambda t_{0} }\,.  
\end{equation} 
Thus, even the weakest measurement (small $\lambda $) turns into
strong a one, if performed often enough for  sufficiently long time.

\section{Representation by stochastic evolution equations}

Turning back to the general case, we note first that the correlation
functions associated with the quasiprobability $P_{q}[a(t)]$, given by
Eqs.~(\ref{av1}-\ref{av3}), can be calculated from the Heisenberg
equations
\begin{equation}
  \label{eq:1}
  \rmd\hat{B}_{\phi }(t)/\rmd t=(\rmi/\hbar)[\hat{H}_{\phi }(t)-\phi(t)\hat{A}_\phi(t),\hat{B}_{\phi }(t)]\;,
\end{equation}
where $\hat{A}$ represents the measured variable while $\hat{B}$ is
any system operator.  In particular
\begin{eqnarray}
\rmd\hat{A}_{\phi }(t)/\rmd t & = & (\rmi/\hbar)[\hat{H}_{\phi }(t),\hat{A}_{\phi }(t)],\nonumber\\
\rmd\hat{H}_{\phi }(t)/\rmd t & = & \phi(t)d\hat{A}_\phi(t)/\rmd t.\label{heiseq}
\end{eqnarray}
We can solve these equations for a general stochastic trajectory
$\phi(t)$, then take the averages as defined by
Eqs.~(\ref{av1}-\ref{av3}), over a Gaussian distribution of such
trajectories. The correlation functions obtained in this way
coincide with the ones derived directly from the CGF Eq.~(\ref{pqq}).  If
$\hat{H}=\hat{p}^{2}/2m +V(\hat{x})$, with
$[\hat{x},\hat{p}]=\rmi\hbar\hat{1}$, and $\hat{A}=\hat{x}$ is the
position operator, the Heisenberg equations for $\hat{x}_{\phi } $ and
$\hat{p}_{\phi } $ are
\begin{eqnarray} \label{langv} 
&&\hat{H}_\phi=\hat{p}^{2}_\phi/2m +V(\hat{x}_\phi),\nonumber\\
&&\frac{\rmd\hat{x}_{\phi } }{\rmd t} =\hat{p}_{\phi }/m, \\ 
&&\frac{\rmd\hat{p}_{\phi } }{\rmd t} =(\rmi/\hbar)[V(\hat{x}_{\phi }),\hat{p}_{\phi }]+\hbar \phi 
(t)=-\frac{\partial V(\hat{x}_{\phi })}{\partial \hat{x}_{\phi } } +\hbar \phi(t).
\nonumber
\end{eqnarray} 
(\ref{langv}) is a quantum Langevin equation in which the quantum
dynamics is augmented by a zero centered white Gaussian noise,
$\langle \phi(t)\rangle =0$, $\langle \phi(t)\phi(t')\rangle =\lambda
\delta(t-t')$. Closed form solutions of this equation can be obtained
for the harmonic oscillator, a case we discuss below. 

Alternatively, the stochastic dynamics affected by the continuous
measurement process may be described by a Lindblad-type master
equation\cite{lind} for the non-selective system density matrix. The
latter is defined by
\begin{equation} \label{23)} 
\hat{\tilde{\rho }}(t)=\int_{a(0)}^{a(t)}  \rmD a\hat{\rho }[a]=\int_{a(0)}^{a(t)} \rmD a\,\hat{K}[a]\hat{\rho }\hat{K}^{\dag }[a]  
\end{equation} 
where the integral is over all observation trajectories between times
$0$ and $t$. It is shown (appendix E) to evolve according to  (using a
Liouville-superoperator $\check L$)
\begin{equation} \label{stoch} 
\frac{\rmd\hat{\tilde{\rho }}}{\rmd t} =\breve{L}\hat{\tilde{\rho }}:=[\hat{H},\hat{\tilde{\rho }}]/\rmi\hbar -\lambda [\hat{A},[\hat{A},\hat{\tilde{\rho }}]]/2.
\end{equation} 
In the representation of eigenstates of $\hat{A}$,  
\begin{equation} \label{25)} 
\hat{\tilde{\rho }}=\sum _{a,a'}\tilde{\rho }_{aa'} {\left| a \right\rangle} {\left\langle a' \right|},
\end{equation} 
\begin{equation} \label{laa} 
\breve{L}\tilde{\rho }_{a,a'} =\frac{1}{\rmi\hbar } \sum _{b}\left(H_{ab} \tilde{\rho }_{ba'} -\tilde{\rho }_{ab} H_{ba'} \right) -\lambda \left(a-a'\right)^{2} \tilde{\rho }_{aa'}  ,
\end{equation} 
showing, as is well known \cite{wimi} and as may be intuitively
expected, that the measurement damps the off-diagonal terms ($a\ne
a'$) with the rate proportional to the measurement strength. Note that
if some eigenvalues $a$ are degenerate, then the corresponding
off-diagonal elements of $\hat{\tilde{\rho }}$ are not damped.

Together with the Liouville-Lindblad superoperator $\breve{L}$ we
define the corresponding evolution superoperator $\breve{U}(a,b)={\rm
  {\mathcal T}}\exp\int _{b}^{a}\breve{L}\rmd t$. It can be then shown
(Appendix F) that the correlation functions (\ref{av1}-\ref{av3})
are given by
\numparts
\begin{eqnarray}
 &&\langle a(t)\rangle _{q} ={\rm Tr}\left[\breve{A}\breve{U}(t,0)\hat{\tilde{\rho }}\right],\label{avu1}\\       %(\label{ZEqnNum751604}a)
&&\langle a(t_1)a(t_2)\rangle _{q} ={\rm Tr}\left[\breve{A}\breve{U}(t_2,t_1)\breve{A}\breve{U}(t_1,0)\hat{\tilde{\rho }}\right],
\label{avu2}\\ %    (\eqref{ZEqnNum751604}b)
&&\langle a(t_1)a(t_2)a(t_3)\rangle _{q} =\label{avu3}
{\rm Tr}\left[\breve{A}\breve{U}(t_3,t_2)\breve{A}\breve{U}(t_2,t_1)\breve{A}\breve{U}(t_1,0)\hat{\tilde{\rho }}\right],\\ %   (\eqref{ZEqnNum751604}c)&&\langle a(t_1)a(t_2)a(t_3)\rangle _{q} =\label{avu3}\\
&&\langle a(t_1)\cdots a(t_n)\rangle _{q} =
{\rm Tr}\left[\breve{A}\breve{U}(t_n,t_{n-1})\cdots\breve{A}\breve{U}(t_2,t_1)\breve{A}\breve{U}(t_1,0)\hat{\tilde{\rho }}\right],\nonumber %   (\eqref{ZEqnNum751604}c)
\end{eqnarray}
\endnumparts
where $\breve{A}\hat{B}=\{ \hat{A},\hat{B}\} /2$. Note that in
(\ref{avu1}-\ref{avu3}), $\hat{\tilde{\rho }}=\hat{\tilde{\rho
  }}\left(t=0\right)=\hat{\rho }$. (\ref{avu1}-\ref{avu3})
provide a more convenient route for the evaluation of these
correlation functions.

In the following Sections we apply this general formalism to the two
simplest quantum systems, the two-level system and the harmonic
oscillator.

\section{The two-level system}

Consider a two-level system defined by the Hamiltonian 
\begin{equation} \label{28)} 
\hat{H}=\hbar \omega \hat{\sigma }_{x} /2 
\end{equation} 
and suppose that the system is in the initial state
\begin{equation} \label{sri} 
\hat{\rho }\left(t=0\right)=(\hat{1}+\hat{\sigma }_{z} )/2,         
\end{equation} 
where $\hat{\sigma }$ denotes Pauli matrices and $\hat{1}$ is the
corresponding unit operator. Left uninterrupted, the system will
oscillate between the two eigenstates of $\hat{\sigma }_{z} $, a
process analogous to Rabi oscillations in a harmonically driven
system. We focus on the measurement of $\hat{A}=\hat{\sigma }_{z} $
and denote the measurement outcome by $a\left(t\right)=\sigma _{z}
\left(t\right)$. We pose the following questions: Can the oscillatory
time trace of $\sigma _{z} $ be observed? How does the measurement
process affect this oscillation? Is the oscillation visible in a
single run of an experiment or only as a statistical effect -- average
over many runs or many copies of the same experiment?  The latter
question is particularly relevant in light of the growth of activity
in single molecule spectroscopy.

To answer these questions we start by writing the action of
$\breve{L}$, (\ref{stoch}), in the basis of Hermitian operators
$(\hat{\sigma }_{x} ,\hat{\sigma }_{y} ,\hat{\sigma }_{z} )$. In a
compact notation it reads,
\begin{equation} \label{lxyz} 
\breve{L}(x\hat{\sigma }_{x} +y\hat{\sigma }_{y} +z\hat{\sigma }_{z} )=\omega (y\hat{\sigma }_{z} -z\hat{\sigma }_{y} )-2\lambda (x\hat{\sigma }_{x} +y\hat{\sigma }_{y} ) 
\end{equation} 
and $\breve{L}\hat{1}=0$. Next, expressing the operation of
$\breve{U}\left(t,0\right)$ on $\hat{\sigma }_{z} $ by
\begin{equation} \label{suu} 
\hat{\sigma }_{z} \left(t\right)=\breve{U}(t,0)\hat{\sigma }_{z} =x(t)\hat{\sigma }_{x} +y(t)\hat{\sigma }_{y} +z(t)\hat{\sigma }_{z}  
\end{equation} 
and using (\ref{lxyz}) and (\ref{suu}) in (\ref{stoch}) we find
${\rmd x\mathord{\left/ {\vphantom {\rmd x \rmd t}}
    \right. \kern-\nulldelimiterspace} \rmd t} =-2\lambda x$;
${\rmd y\mathord{\left/ {\vphantom {\rmd y \rmd t}}
    \right. \kern-\nulldelimiterspace} \rmd t} =-\left(\omega z+2\lambda
  y\right)$, and  ${\rmd z\mathord{\left/ {\vphantom {\rmd z \rmd t}}
    \right. \kern-\nulldelimiterspace} \rmd t} =\omega y$, which, for
$z(t=0)=1$, $x(0)=y(0)=0$ yields 
\begin{equation} \label{ztt} 
z(t)=\rme^{-\lambda t} [\cos (\Omega t)+\lambda \sin (\Omega t)/\Omega ],\;
y(t)=\rmd z/\rmd t,\; x(t)= 0,
\end{equation} 
where $\Omega =\sqrt{\omega ^{2} -\lambda ^{2} } $. This allows to
write down the relevant averages (see Appendix G), namely
\begin{equation} \label{avs12} 
\langle \sigma _{z} (t)\rangle _{q} =z(t),\; 
\langle \sigma _{z} (t)\sigma _{z} (t')\rangle _{q} =z(|t-t'|).
\end{equation} 
The last line is known in the existing literature only for the
stationary case ($t,t'\to \infty$) in the weak measurement limit
$\lambda\to 0$ \cite{paz,kor2}.  It is interesting to note that
although the system under consideration is not in a stationary state
and in fact evolves irreversibly, this correlation function depends
only on the time difference $t'-t$ and remains finite when this
difference is constant while both $t$ and $t'$ increase.

Recall that (\ref{aaa}) and (\ref{acor}) imply that $\langle
\sigma _{z} (t)\rangle =\langle \sigma _{z} (t)\rangle _{q} $, while
$\langle \sigma _{z} (t)\sigma _{z} (t')\rangle =\langle \sigma _{z}
(t)\sigma _{z} (t')\rangle _{d} +\langle \sigma _{z} (t)\sigma _{z}
(t')\rangle _{q} $, which implies $\langle \delta \sigma _{z}
(t)\delta \sigma _{z} (t')\rangle $ $=\langle \delta \sigma _{z}
(t)\delta \sigma _{z} (t')\rangle _{d} +\langle \delta \sigma _{z}
(t)\delta \sigma _{z} (t')\rangle _{q} $. In the limit
$\lambda\ll\omega $ we see (cf. (\ref{ztt}) and (\ref{avs12}))
clear oscillation of $\langle \sigma _{z} (t)\rangle $. However, in a
single run this signal cannot be distinguished from the noise. Indeed,
defining as in (\ref{ava}) $\bar{\sigma }_{z} =\left(1/t_{0}
\right)\int _{0}^{t_{0} } \rmd t\sigma _{z} (t)$, we obviously need to
take $t_{0}\ll\omega ^{-1} $. Therefore
\begin{equation} \label{stn} 
\langle (\delta \bar{\sigma }_{z} )^{2} \rangle >\langle ((\delta \bar{\sigma }_{z} )^{2} \rangle _{d} =1/t_{0} \lambda \gg 1 
\end{equation} 
The large detection noise covers the signal. This implies that Rabi
oscillations cannot be seen in a single run.

The above result was obtained in the time domain. We can also ask
whether the Rabi oscillation is visible in the frequency domain. This
would imply seeing the peak in the Fourier transform
\begin{equation} \label{sfreq} 
\tilde{\sigma }_{z} \left(\nu \right)=(2/t_0)\int _{0}^{t_{0} } \rmd t\cos (\nu t)\sigma _{z} (t) ,      
\end{equation} 
where $t_{0} $ is a time much larger than the oscillation period, but
obviously much smaller than the damping time: $\omega ^{-1}\ll
t_{0}\ll\lambda ^{-1} $. From (\ref{ztt}) and (\ref{avs12}), the peak
intensity is $\langle\tilde{\sigma }_{z} \left(\nu=\Omega
\right)\rangle\simeq 1$. On the other hand, under the measurement
conditions the white detector noise satisfies
\begin{equation} \label{36)} 
\langle (\delta \tilde{\sigma}_z(\Omega))^{2} \rangle\geq\langle
(\delta \tilde{\sigma}_z(\Omega))^{2} \rangle _{d} =1/2t_{0} \lambda
\gg 1  
\end{equation} 
implying that, again, the noise exceeds the signal and a peak in the
frequency domain will not be seen. This time, however, the
signal-noise ratio is not as bad as in the time domain because $t_{0}
$ can be longer.

We conclude that Rabi oscillation cannot be seen in a single run/copy
of the experiment but only in a statistical average. The sample size,
that is the number of runs/copies needed for this average is of the
order $(t_{0} \lambda )^{-1} $, where $\lambda ^{-1}\gg t_{0}\gg\omega
^{-1} $ in the frequency domain and $t_{0} \ll \omega ^{-1}\ll\lambda
^{-1} $ in the time domain. In the overdamped regime,
$\lambda\gg\omega $, one can see the quantum Zeno effect, discussed
below in Section 7.

\section{Leggett-Garg-type inequality}

\begin{figure}
\includegraphics[scale=1,angle=270]{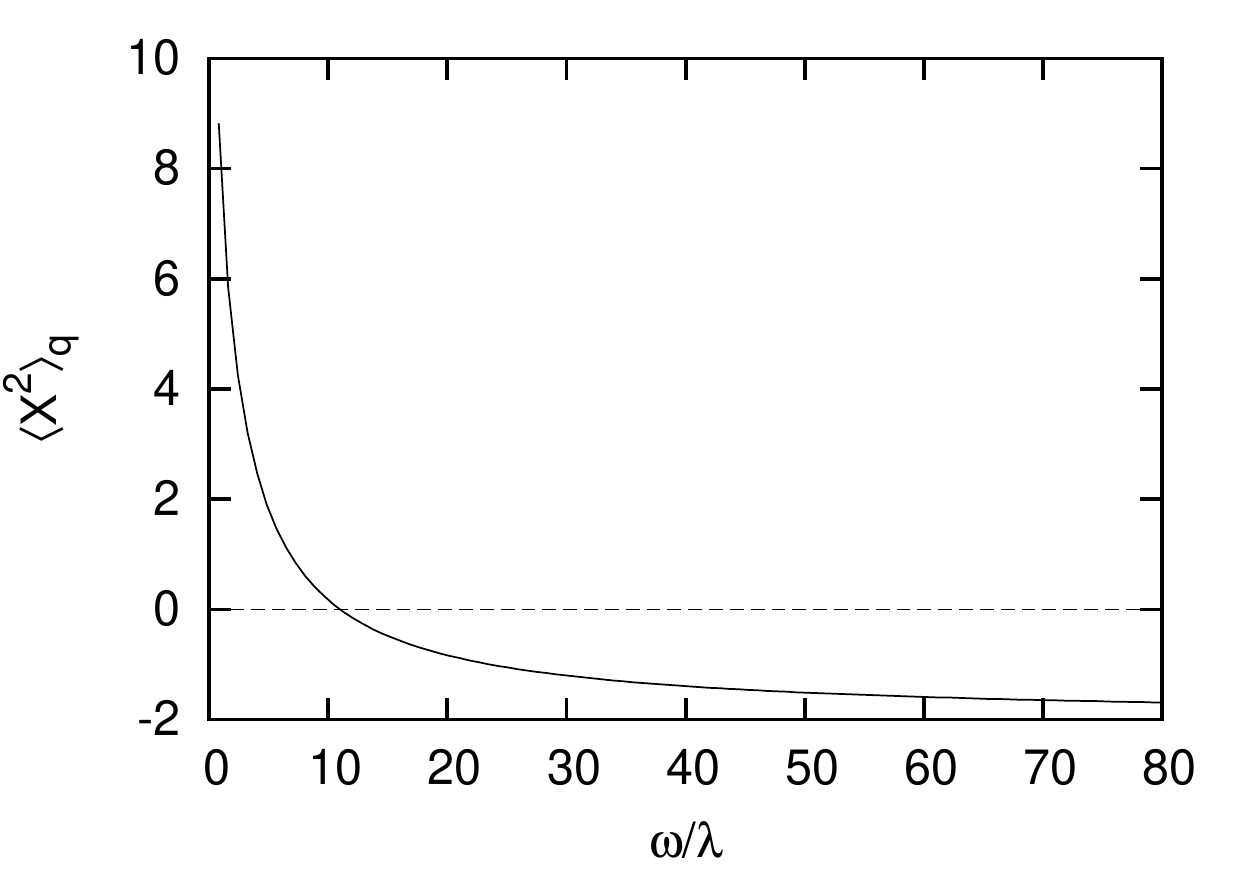}
\caption{Demonstration of the violation of the Leggett-Garg-type
  inequality $\langle X^2\rangle_q>0$ as a function of
  measurement strength $\lambda$. The function starts from the value
  $16$ for $\lambda\gg\omega$ as expected for the QZE, crosses the classical
  bound $0$ and tends to $-2$ in the limit of weak measurement
  $\lambda\ll \omega$.  }\label{xlam2}
\end{figure}

The limit $\lambda\to 0$ is consistent with the noninvasive measurement
because the backaction vanishes.
In this case the negative quasiprobability demonstrates the violation of macrorealism
even for a single observable,
as shown by Leggett and Garg \cite{leg}. In violations of this type, it is essential to subtract 
the large detection noise, whose uncertainty must always diverge and prevent any real violation \cite{kobr}.
The common confusion about the noninvasiveness condition is caused by the fact that
two-time correlations are numerically identical for the quasiprobability in the
limit $\lambda\to 0$ and the instant projections (invasive because of collapse) for initial $\hat{\rho}\sim\hat{1}$.
The equality still holds in the case of many times if the observable satisfies $\hat{A}^2\propto \hat{1}$.
The analysis above has used second order correlations that, as stated
in (\ref{wepo}), are not sensitive to the quasiprobabilistic nature of
the distribution. The violation of the well known Leggett-Garg inequality \cite{leg}
needs only second order correlations but requires the additional
assumption of bounded observables which is effectively equivalent to
higher order correlations (e.g. the dichotomy $A=\pm 1$ is equivalent
to measuring $\langle(A^2-1)^2\rangle=0$ which requires the
fourth-order correlator $\langle A^4\rangle$).  Without this
assumption, the quasiprobabilistic nature is however revealed in
higher order correlations. To see this we take $\hat{\rho}=\hat{1}/2$
and consider the following quantity
\begin{equation}
X[\sigma]=\sigma_z(0)\sigma_z(\pi/\omega)+\sigma_z(-\pi/2\omega)\sigma_z(\pi/2\omega)+2\label{xxxa}.
\end{equation}
The fourth order correlation $\langle X^2\rangle_q$ is given by
\begin{equation}
\langle X^2\rangle_q=
6+\label{xtau}
\rme^{-\lambda\pi/\omega}[1/r^2+(10-1/r^2)\cos(\pi r)+10\lambda\sin(\pi r)/\omega r]
,
\end{equation}  
where $r=\sqrt{1-(\lambda/\omega)^2}$. The behavior of $\langle
X^2\rangle_q$ is shown in figure \ref{xlam2}. In the limit of strong
measurement $\langle X^2\rangle_q=16$. The origin is the QZE -- the
evolution is frozen by the measurement and so $\sigma_z(t)$ does not
depend on time, which results in $X=4$. In the opposite limit of
noninvasive measurement $\langle X^2\rangle_q=-2$ and it crosses zero
at $\omega/\lambda\approx 11$.  This implies that for sufficiently
small $\lambda$ the classical inequality $\langle X^2\rangle_q\geq 0$
is violated so the function $P_q$ is not positive definite and as such
cannot describe a usual probability.  Note, however, that (a) it contains
the relevant physical information, discarding the irrelevant detection noise; 
(b) by itself, it cannot be directly measured, namely
correlations such as $\langle X^2\rangle_q$ are not directly
measurable since the real probability is the convolution (\ref{conv});
and (c) the actual detected observable certainly satisfies $\langle
X^2\rangle>0$. However, an independent determination of the detection
noise should be experimentally feasible and allows to find the
negativity of $\langle X^2\rangle_q$ after the noise has been
subtracted.

\section{The harmonic oscillator}
For completeness, we also consider another much
studied simple problem -- continuous position measurement,
$\hat{A}=\hat{x}$, in a system comprising one harmonic oscillator \cite{gpovm},
described by the Hamiltonian $\hat{H}=\hat{p}^{2} /2m+m\omega ^{2}
\hat{x}^{2} /2$. Eqs.~(\ref{langv}) become
\begin{equation} \label{xph} 
\begin{array}{l} {\rmd\hat{x}_{\phi } /\rmd t=\hat{p}_{\phi } /m,} \\ 
{\rmd\hat{p}_{\phi } /\rmd t=-m\omega ^{2} \hat{x}_{\phi } +\hbar \phi (t),} \end{array} 
\end{equation} 
where $\phi $ represents the zero-centered white Gaussian noise,
$\langle \phi (t)\phi (t')\rangle =\lambda \delta (t-t')$. We note in
passing that this quantum Langevin equation yields the Fokker-Planck
equation for the Wigner function \cite{wigner}
\begin{equation} \label{38)}
W\left(x,p\right)=\int \frac{\rmd\chi \rmd\xi }{(2\pi )^{2} }  
\rme^{-\rmi\xi x-\rmi\chi p} {\rm Tr}\hat{\rho }\rme^{\rmi\xi x+\rmi\chi p}  
\end{equation} 
 in the form \cite{wfp}
\begin{equation} \label{39)} 
\frac{\partial W\left(x,p,t\right)}{\partial t} =m\omega ^{2} x\frac{\partial W}{\partial p} -\frac{p}{m} \frac{\partial W}{\partial x} +\frac{\lambda \hbar ^{2} }{2} \frac{\partial ^{2} W}{\partial p^{2} }.
\end{equation} 
However, in what follows we calculate directly the required
correlation functions. Solving (\ref{xph}) we get
\begin{eqnarray}
&&\hat{x}_\phi(t)=\hat{x}(0)\cos \omega t+\frac{\hat{p}(0)}{m\omega}\sin \omega t+\int _{0}^{t} \frac{\rmd t'}{m\omega}\; \sin \omega (t-t')\; \hbar \phi (t) ,
\label{xph1}\\
 %(\label{ZEqnNum752357}a)
&&\hat{p}_\phi(t)=\hat{p}(0)\cos \omega t-m\omega \hat{x}(0)\sin \omega t+\int _{0}^{t} \rmd t'\; \cos \omega (t-t')\; \hbar \phi (t'),\nonumber
% (\eqref{ZEqnNum752357}b)
\end{eqnarray}
This implies that the oscillations of the average position are
undamped,
\begin{equation} \label{avx} 
\left\langle x\left(t\right)\right\rangle =\left\langle x\left(0\right)\right\rangle \cos \left(\omega t\right)+\left(m\omega \right)^{-1} \left\langle p\left(0\right)\right\rangle \sin \left(\omega t\right) 
\end{equation} 
independently of the detection strength. 

Turning to the noise term we first note that, as before, the detector
noise combines additively with the the correlation functions
obtained from (\ref{xph1}). The latter take the form
\begin{equation} \label{42)} 
\left\langle \delta x\left(t\right)\delta x\left(t'\right)\right\rangle _{q} =\left\langle \delta x\left(t\right)\delta x\left(t'\right)\right\rangle _{0} +f_{\lambda } \left(t,t'\right) 
\end{equation} 
where $\langle \delta x(t)\delta x(t')\rangle _{0} $ is the free
correlation function obtained in the limit $\lambda \to 0$ or
equivalently $\phi \to 0$, that is, by ignoring the last (noise) terms
on the RHS of (\ref{xph1}),
\begin{eqnarray}
&&\langle \delta x(t)\delta x(t')\rangle _{0} =\langle \delta x(0)\delta x(0)\rangle \cos \omega t\cos \omega t'+ \label{43)}\\
&&\langle \delta x(0)\delta p(0)\rangle _{W} (m\omega )^{-1} \sin \omega (t+t') 
+\langle \delta p(0)\delta p(0)\rangle (m\omega )^{-2} \sin \omega
t\sin \omega t'\nonumber
\end{eqnarray} with the Wigner-ordered average $\langle 2xp\rangle
_{W} ={\rm Tr}\left[\hat{\rho }\{ \hat{x},\hat{p}\} \right]$, and
where $f_{\lambda } \left(t,t'\right)$ is the correlation function
associated with the noise terms in (\ref{xph1}),
\begin{eqnarray} \label{ffl} 
&&f_{\lambda } (t,t')=\\
&&\frac{\lambda \hbar ^{2} }{2(m\omega )^{2} }[\min (t,t')\cos \omega (t-t')+(\sin \omega |t-t'|-\sin \omega (t+t'))/2\omega ].\nonumber
\end{eqnarray} 
This measurement-induced correlation function represents the
backaction effect of the measuring process. It depends on the detector
strength and the parameters of the dynamics but not on the initial
state of the oscillator. Moreover, because of the Gaussian nature of
$\phi $, it contributes solely to the second cumulant $\langle \langle
x(t)x(t')\rangle \rangle $, leaving all the other unaffected. As
expected, it vanishes in the limit $\lambda \to 0$. However, the most
striking feature in (\ref{ffl}) is the growth of noise with time, as
expressed by the first term in (\ref{ffl}).

In analogy to the two-level system we discuss the behavior of the
short time ($t\ll \omega ^{-1} $) average $\bar{x}$ and the long time
($t\gg \omega ^{-1} $) Fourier transform $\tilde{x}$, defined by the
analogs of (\ref{ava}) and (\ref{sfreq}), respectively. In both
limits we are now free to choose $t_{0} \lambda $ because, in contrast
to the 2-level case, the averaged oscillations, (\ref{avx}), are
not damped.

Consider first the time-domain observation. For $t_{0} \lambda\ll 1$
the uncertainty of $\bar{x}$ is determined by the detection noise as
in (\ref{stn}), while in the opposite limit it will be dominated by
the backaction (\ref{ffl}). In either case the noise exceeds the
signal.

In the frequency domain, for $\tilde{x}\left(\omega \right)=2\int
_{0}^{t_{0} } \rmd t\cos (\omega t)x(t)/t_{0} $, we get using
(\ref{avx}-\ref{ffl}), the peak signal
\begin{equation}
\left\langle \tilde{x}\left(\omega \right)\right\rangle =\left\langle
  x(0)\right\rangle , \label{xfreq} 
\end{equation}
and the intrinsic and backaction noise components
\begin{equation}
 \left\langle \left(\delta \tilde{x}\left(\omega \right)\right)^{2}
 \right\rangle _{q} \simeq \left\langle \left(\delta
     x\left(0\right)\right)^{2} \right\rangle +\frac{\lambda t_{0}
   \hbar ^{2} }{6\left(m\omega \right)^{2}}\label{xxfreq} 
 \end{equation}
 to which we need to add the detector noise $(2t_{0} \lambda)^{-1}
 $. The total uncertainty originated from the detector satisfies
\begin{equation} \label{46)} 
\frac{\lambda t_{0} \hbar ^{2} }{6\left(m\omega \right)^{2} }
+\frac{1}{2t_{0} \lambda } \ge \frac{\hbar }{\sqrt{3} m\omega } 
\end{equation} 
with the lower bound (obtained as minimum of the LHS with respect to $t_{0}
\lambda $) independent of $\lambda $ and $t_{0} $. Obviously
$\left\langle x\left(0\right)\right\rangle $ can be chosen large
enough for the signal to dominate the noise at intermediate times, but
the noise will always exceed the signal at long enough times. As
always, the signal-to-noise ratio can be improved by repeated
measurements.

The fact that the backaction contribution (\ref{ffl}) to the noise
grows with time reflects the continuous pumping of energy to the
system affected by the measurement process \cite{sem}. This does not happen in
the two-level system because of its bounded spectrum, still also in
that system the temperature grow to infinity ($\hat{\rho }(t) \to
(1/2)\hat{1}$) as implied by Eqs.~(\ref{suu}), (\ref{ztt}) and
(\ref{avs12}). This unlimited growth can be avoided by
assuming that the measurement process also involves some friction
\cite{cav2,wfp}. Indeed, measurement, even classical, means extraction
of information out of the system, so that without compensating friction its
entropy must increase and so does the temperature.

\section{The quantum Zeno effect (QZE)}

For completeness, we show now how the QZE emerges within the present formalism. 
So far we have focused on weak measurements, represented by small
$\lambda $. The opposite limit, $\lambda \to \infty $, represents the
strong measurement case. In systems characterized by a single timescale
$\omega ^{-1} $, strong and weak measurements are quantified by the
inequalities $\lambda\gg\omega $ and $\lambda \ll \omega $,
respectively.

Consider the two-level system discussed in Section 4. For $\lambda
>\omega $ its dynamics is given by the overdamped analog of
Eq.~(\ref{ztt}), $\Omega =\rmi\sqrt{\lambda ^{2} -\omega ^{2} } $.  In
the extreme strong measurement case, $\lambda\gg\omega $, $z(t)\sim
\rme^{-\omega ^{2} t/2\lambda } $ and the decay slows down as $\lambda
\to \infty $\cite{gag, schul}. This corresponds to the QZE where the
system is almost frozen by the measurement, reaching its equilibrium
state $z=0$ only on the timescale $t\sim\lambda /\omega ^{2} $.

For a position measurement in the harmonic oscillator case, we have
seen, Eq.~(\ref{avx}), that the average position oscillates regardless
the strength of the measurement. This implies that the Zeno affect is
absent in this system, as is well known \cite{gag}. On the other hand,
for any measurement strength, the detector-induced backaction noise,
Eq.~(\ref{ffl}), increases without bound at long times at a rate that
increases with \textit{$\lambda $}. Already for short times we get
$f_{\lambda } (t,t)\simeq\lambda \hbar ^{2} /3m^{2} $, and backaction
adds fast diffusion in the phase space. This is somewhat analogous to
the anti-Zeno effect \cite{aze}.

\section{Conclusions}

Gaussian POVMs, here represented by Kraus operators, were used in this
paper to formalize the description of weak measurements. A path
integral representation of continuous weak measurement described in
this way leads directly to an analysis of backaction noise in terms of
stochastic evolution equations. The average signal and the associated
noise were obtained in terms of moments and time
correlation functions of the measured quantity. 

In particular, the noise was shown to be an additive combination of a
term characteristic of the measurement alone (detector noise) and
terms associated with the system, which in turn include contributions
from the intrinsic quantum mechanical uncertainty in the system and
from backaction effects from the measurement process.  A transparent
representation of this stochastic evolution was obtained by separating
it into a process characteristic only of the weak measurement, and
another, representing the quantum uncertainty intrinsic to the system
as well as that arising from the measurement backaction. This yields
the noise as an additive combination of the corresponding
contributions, while the total probability is found to be convolution
of white Gaussian detections noise and intrinsic system's
quasiprobability.  The quasiprobability can be negative although the
negativity is not visible at the level of second order correlations
due to weak positivity.  The general formalism was applied to two
simple problems: continuous monitoring of the level population in a
2-level system and continuous measurement of the position of a
harmonic oscillator. For both systems we have established limits on
the possibility to observe oscillatory motion in a single run of an
experiment. The negativity property of the quasiprobability can be
demonstrated on in the 2-level system, using fourth order
correlations. In this way, we have constructed a Leggett-Garg-type
inequality without the assumption of dichotomy or boundedness of the variable.

We observe that the QZE occurs when both the Hamiltonian and the
observable can be represented in finite-dimensional Hilbert
space. When the space is infinite or continuous and both the
Hamiltonian and the observable have no finite-dimensional
representation, the dynamics will not always be able to 'pin down' the
state and consequently the dynamics may get diffusive.  Establishing
criteria for the occurrence or absence of the QZE in realistic systems
continues to be an intriguing and challenging issue.

\section*{Acknowledgments}
We are grateful to J. Audretsch and Y. Aharonov for fruitful
discussions.  AN acknowledges support by the European Research Council
(FP7/2007-2013 grant agreement no. 226628), the Israel - Niedersachsen Research Fund,
and the Israel Science Foundation. He also thanks the Alexander von
Humboldt Foundation and the SFB 767 for sponsoring his visit at the
University of Konstanz. WB and AB acknowledge support by the DFG
through SFB 767 and SP 1285. All authors also wish to thank the Kurt
Lion Foundation for supporting this work.

\section*{Appendix A} 
\renewcommand{\theequation}{A.\arabic{equation}}
\setcounter{equation}{0}

Here we derive (\ref{kra2}).
Using (\ref{kra1}), a succession of time evolutions and measurements in the interval $\left(0,t\right)$ reads
\begin{eqnarray} \label{47)} 
&&\hat{K}\left(\left\{a_{j} \right\}\right)=(2\bar{\lambda }/\pi )^{N/4}\rme^{-\left(\rmi/\hbar \right)\hat{H}\left(t_{N+1} -t_{N} \right)} \rme^{-\bar{\lambda }(a_{N} -\hat{A})^{2} }
\cdots \times\nonumber\\
&&\rme^{-\left(\rmi/\hbar \right)\hat{H}\left(t_{3} -t_{2} \right)} \rme^{-\bar{\lambda }(a_{2} -\hat{A})^{2} }\rme^{-\left(\rmi/\hbar \right)\hat{H}\left(t_{2} -t_{1} \right)} \rme^{-\bar{\lambda }(a_{1} -\hat{A})^{2} } \rme^{-\left(\rmi/\hbar \right)\hat{H}t_{1} }.
\end{eqnarray} 
Putting $\bar{\lambda }=\lambda \Delta t$ and $t_{j} -t_{j-1} =t_{1} =\Delta t$ and using $\Delta t\to 0$ leads to
\begin{equation}
\hat{K}\left(\left\{a_{j} \right\}\right)=(2\lambda\Delta t /\pi )^{N/4}
\prod _{j=1}^{N}\rme^{\left[-\left(\rmi/\hbar \right)\hat{H}-\lambda \left(a_{j} -\hat{A}\right)^{2} \right]\Delta t}  \rme^{-\left(i/\hbar \right)H\Delta t}
\end{equation}
and, for $\Delta t\to 0$
\begin{equation}\label{kk1}
\hat{K}\left[a\left(t\right)\right]=C{\mathcal T}\rme^{\int _{0}^{t}\left[-\left(\rmi/\hbar \right)\hat{H}-\lambda \left(a\left(t\right)-\hat{A}\right)^{2} \right]\rmd t }.
\end{equation}
 Alternatively, using $\hat{A}\left(t_{j} \right)=\rme^{\left(\rmi/\hbar \right)\hat{H}t_{j} }\hat{A}
\rme^{-\left(\rmi/\hbar \right)\hat{H}t_{j} } $ yields
\begin{equation}
\hat{K}\left(\left\{a_{j} \right\}\right)=(2\lambda\Delta t /\pi )^{N/4}
\rme^{-\left(\rmi/\hbar \right)\hat{H}t_{N+1} } \prod _{j=1}^{N}\rme^{\left[-\lambda \left(a_{j} -\hat{A}\left(t_{j} \right)\right)^{2} \right]\Delta t} 
\end{equation}
and in the continuum limit
\begin{equation}\label{kk2}
K\left[a\left(t\right)\right]=Ce^{-\left(\rmi/\hbar \right)Ht} {\mathcal T}\rme^{-\lambda \int _{0}^{t}\left(a\left(t\right)-\hat{A}\left(t\right)\right)^{2} \rmd t }
\end{equation}
 In (\ref{kk1}) and (\ref{kk2}) $C$ are normalization factors.

\section*{Appendix B} 
\renewcommand{\theequation}{B.\arabic{equation}}
\setcounter{equation}{0}

Here we prove (\ref{hphi}).
Start from ${\mathcal T}\exp\int  \rmi(\chi (t)/2+\phi (t))\hat{A}(t)\rmd t$ and discretize to get
\begin{eqnarray}
&&{\mathcal T}\rme^{\rmi\int (\chi (t)/2+\phi (t))\hat{A}(t)\rmd t } =
{\mathcal T}\rme^{\rmi\Delta t\sum _{j}(\chi (t_{j} )/2+\phi (t_{j} ))\hat{A}(t_{j} ) }
= \left\{t_{j} =j\Delta t;\, j=1,...N\right\}\nonumber \\
 &&\rme^{\rmi\Delta t(\chi (t_{N} )/2+\phi (t_{N} ))\hat{A}(t_{N} )}\rme^{\rmi\Delta t(\chi (t_{N-1} )/2+\phi (t_{N-1} ))\hat{A}(t_{N-1} )}
 \cdots \rme^{\rmi\Delta t(\chi (t_{1} )/2+\phi (t_{1} ))\hat{A}(t_{1} )}
\nonumber \\ 
&&=\rme^{\left(\rmi/\hbar \right)Ht_{N} } \rme^{\rmi\Delta t(\chi (t_{N} )/2+\phi (t_{N} ))\hat{A}}\rme^{-\left(rmi/\hbar \right)Ht_{N} }\times\nonumber\\ 
&&\rme^{\left(\rmi/\hbar \right)Ht_{N-1} }\rme^{\rmi\Delta t(\chi (t_{N-1} )/2+\phi (t_{N-1} ))\hat{A}} \rme^{-\left(\rmi/\hbar \right)Ht_{N-1} } \cdots
 \nonumber\\
&&\cdots\, \rme^{\left(\rmi/\hbar \right)Ht_{1} } \rme^{\rmi\Delta t(\chi (t_{1} )/2+\phi (t_{1} ))\hat{A}} \rme^{-\left(\rmi/\hbar \right)Ht_{1} } \end{eqnarray}
Next replace
\begin{equation}
\rme^{\rmi\Delta t(\chi (t_{j} )/2+\phi (t_{j} ))\hat{A}} \to \rme^{\rmi\Delta t\hat{A}\chi (t_{j} )/2} \rme^{\rmi\hat{A}\phi (t_{j} )\left(t_{j} -t_{j-1} \right)}
\end{equation}
for  $j=1,...N$
and define $\hat{H}_{\phi } \left(t\right)=\hat{H}-\hbar \phi \left(t\right)\hat{A}$,\newline $\hat{A}_{\phi }(t)=
\tilde{\mathcal T} \rme^{\left(\rmi/\hbar \right)\int 0^t\hat{H}_{\phi }(t')\rmd t' } \hat{A}\mathcal T
\rme^{-\left(\rmi/\hbar \right)\int_0^t\hat{H}_{\phi } \left(t'\right)\rmd t' } $ 
and discretize it again,
\begin{equation}
\hat{A}_{\phi }(t_k)=
\tilde{\mathcal T}\prod_{j\leq k} \rme^{\left(\rmi/\hbar \right)\hat{H}_{\phi }(t_j)\Delta t } \hat{A}\mathcal T
\prod_{j\leq k}\rme^{-\left(\rmi/\hbar \right)\hat{H}_{\phi }(t_j)\Delta t }
\end{equation}
to get
\begin{equation}
\rme^{\rmi t_{N} \phi \left(t_{N} \right)} {\mathcal T}\rme^{\rmi\Delta t\sum _{j=1}^{N}\chi(t_j)\hat{A}_{\phi }(t_j)/2 } \to \rme^{\rmi t\phi \left(t\right)} {\mathcal T}\rme^{\rmi\int_0^t (\chi (t')/2)\hat{A}_{\phi } (t')\rmd t' } ,
\end{equation}
from which (\ref{hphi}) follows.

\section*{Appendix C}
\renewcommand{\theequation}{C.\arabic{equation}}
\setcounter{equation}{0}
 Here we derive Eq.(\ref{av1}-\ref{av3}).
 Start from (\ref{sqq}) and take its functional derivatives
\begin{eqnarray} \label{50)} 
&&\left\langle a\left(t\right)\right\rangle _{b} =
\int \rmD a{\kern 1pt} \, a\left(t\right)P_{b} [a] =\frac{1}{i} \left(\frac{\delta \rme^{S_{b}\left[\chi \right]} }{\delta \chi \left(t\right)} \right)_{\chi \left(t\right)=0}\nonumber \\
&&=\int  \rmD\phi {\kern 1pt} \rme^{-\int  \rmd t\phi ^{2} (t)/2\lambda } {\rm Tr}\; {\mathcal T}\left(\hat{A}_{\phi } (t)\hat{\rho }+\hat{\rho }\hat{A}_{\phi } (t)\right)/2\nonumber\\
&&
=\int  \rmD\phi {\kern 1pt} \rme^{-\int  \rmd t\phi^{2} (t)/2\lambda } {\rm Tr}\left(\hat{A}_{\phi } (t)\hat{\rho }\right)
 \end{eqnarray} 
which is (\ref{av1}). 
\begin{eqnarray} \label{corr} 
&&\left\langle a\left(t\right)a\left(t'\right)\right\rangle _{b} =\int \rmD a{\kern 1pt} \, a\left(t\right)a\left(t'\right)P_{b} [a] =
-\left(\frac{\delta^{2} \rme^{S_{b}\left[\chi \right]} }{\delta \chi \left(t\right)\delta \chi \left(t'\right)} \right)_{\chi \left(t\right)=0}
\nonumber\\
&&=
\frac{1}{4} \int  \rmD\phi {\kern 1pt} \rme^{-\int  \rmd t\phi ^{2} (t)/2\lambda } {\rm Tr}\; \left\{\hat{A}_{\phi } (t'),\left\{\hat{A}_{\phi } (t),\hat{\rho }\right\}\right\}
\end{eqnarray} 

Time ordering implies that for $t'>t$, $t'$ will be placed in the outer commutator, however last expression is equal to
\begin{equation} \label{52)} 
\frac{1}{2} \int  \rmD\phi {\kern 1pt} \rme^{-\int  \rmd t\phi ^{2} (t)/2\lambda } {\rm Tr}\; \left\{\hat{A}_{\phi } (t),\hat{A}_{\phi } (t')\right\}\hat{\rho }, 
\end{equation} 
which does not depend on the operator ordering. Higher moments are obtained in the same way. 

\section*{Appendix D}
\renewcommand{\theequation}{D.\arabic{equation}}
\setcounter{equation}{0}
Here we consider the case $[\hat{A},\hat{H}]=0$ and $[\hat{\rho},\hat{A}]=0$.
When the observable $\hat{A}$ commutes with the Hamiltonian, $\hat{A}_{\phi } \left(t\right)=\hat{A}$ so the trace in (\ref{sqq}) becomes independent of \textit{$\phi$}. Using $\int  \rmD\phi {\kern 1pt} \rme^{-\int \rmd t\phi ^{2} (t)/2\lambda  } =1$ it follows that $\rme^{S_{b} [\chi ]} $ and $P_{b} [a]$, (\ref{pqq}), do not depend on $\lambda $. This implies that the evolution associated with the backaction effect is deterministic and the only source of noise is the detector. To see the implication of this on the moments consider the moment generating function (c.f. (\ref{sdd}), (\ref{sqq}))
\begin{eqnarray} \label{53)} 
&&\rme^{S[\chi ]} =\rme^{S_{d} \left[\chi \right]} \rme^{S_{b} \left[\chi \right]} =\rme^{-\int \rmd t\chi ^{2} (t)/8\lambda  } \int  \rmD\phi {\kern 1pt}
 \rme^{-\int \rmd t\phi ^{2} (t)/2\lambda  }\times\nonumber\\
&&{\rm Tr}\; {\mathcal T}\rme^{\rmi\int \chi (t)\hat{A}_{\phi } (t)\rmd t/2 } \; \hat{\rho }\; \tilde{{\mathcal T}}\rme^{\rmi\int \chi (t)\hat{A}_{\phi } (t)\rmd t/2 }
\end{eqnarray} 
For the imposed initial conditions this becomes
\begin{equation} \label{54)} 
\rme^{S[\chi ]} \, =\rme^{-\int \rmd t\chi ^{2} (t)/8\lambda  } \; 
\left(\rme^{\rmi a\int \chi (t)\rmd t/2 } \right)^{2}  
\end{equation} 
The first moment satisfies
\begin{equation} \label{aaa} 
\left\langle a\right\rangle =\left(\frac{\delta }{\rmi\delta \chi \left(t\right)} \rme^{S_{d} [\chi ]+S_{b} [\chi ]} \right)_{\chi \left(t\right)=0}
=\left(\frac{\delta }{\rmi\delta \chi \left(t\right)} \rme^{S_{b} [\chi ]} \right)_{\chi \left(t\right)=0} =a_{0}  
\end{equation} 
which implies also $\left\langle \bar{a}\right\rangle =a $ when used in (\ref{ava}). It is easy to realize that the second moment satisfies 
\begin{eqnarray} \label{acor} 
&&
\left\langle a\left(t\right)a\left(t'\right)\right\rangle =-\left(\frac{\delta ^{2} }{\delta \chi \left(t\right)\delta \chi \left(t'\right)} \rme^{S_{d} [\chi ]+S_{b} [\chi ]} \right)_{\chi \left(t\right)=0}
=\nonumber\\
&&-\left(\frac{\delta ^{2} }{\delta \chi \left(t\right)\delta \chi \left(t'\right)} \rme^{S_{d} [\chi ]} \right)_{\chi \left(t\right)=0}
 -\left(\frac{\delta ^{2} }{\delta \chi \left(t\right)\delta \chi \left(t'\right)} \rme^{S_{b} [\chi ]} \right)_{\chi \left(t\right)=0} \end{eqnarray} 
The second term yields \textit{a}${}^{2}$, so $\left\langle \delta a\left(t\right)\delta a\left(t'\right)\right\rangle =\left\langle \delta a\left(t\right)\delta a\left(t'\right)\right\rangle -a^{2} $ is determined just by the Gaussian detector noise that in
the case of (\ref{ava}) results in
\begin{equation} \label{57)} 
\left\langle (\delta \bar{a})^2\right\rangle =1/\left(4\lambda t_{0} \right).        
\end{equation} 

\section*{Appendix E} 
\renewcommand{\theequation}{E.\arabic{equation}}
\setcounter{equation}{0}

Here we derive the master equation (\ref{stoch}).
Following the steps that lead to (\ref{prob2}) but without the trace, we find
\begin{eqnarray} \label{58)} 
&&\hat{\rho }[a]=\rme^{-\left(\rmi/\hbar \right)\hat{H}t} \int  \rmD\phi {\kern 1pt} \rme^{-\int \rmd t\phi ^{2} (t)/2\lambda  } \int \rmD\chi {\kern 1pt}
\rme^{-\int \rmd t\chi ^{2} (t)/8\lambda  } \rme^{-\int \rmi\chi (t)a(t)\rmd t }\times\nonumber\\
&&  {\mathcal T}\rme^{\int \rmi(\chi (t)/2+\phi (t))\hat{A}(t)\rmd t }
\; \hat{\rho }\; \tilde{{\mathcal T}}\rme^{\int \rmi(\chi (t)/2-\phi (t))\hat{A}(t)\rmd t }
\rme^{\left(\rmi/\hbar \right)\hat{H}t} 
 \end{eqnarray}  
and
\begin{eqnarray} \label{59)} 
&&\hat{\tilde{\rho }}\left(t\right)\equiv \int  \rmD a\hat{\rho }[a]=\nonumber\\
&&\rme^{-\left(\rmi/\hbar \right)\hat{H}t}
\int  \rmD\phi {\kern 1pt} \rme^{-\int \rmd t\phi ^{2} (t)/2\lambda }
{\mathcal T}\rme^{\rmi\int \phi (t)\hat{A}(t)\rmd t } \; \hat{\rho }\; \tilde{{\rm {\mathcal T}}}\rme^{-\rmi\int \phi (t)\hat{A}(t)\rmd t } \rme^{\left(\rmi/\hbar \right)\hat{H}t}  
\end{eqnarray} 
In what follows we will use the incremental propagation version of this equation:
\begin{eqnarray} \label{rrr} 
&&\hat{\tilde{\rho }}\left(t+\Delta t\right)=\rme^{-\left(\rmi/\hbar \right)\hat{H}\left(t+\Delta t\right)} \int  \rmD\phi {\kern 1pt} \rme^{-\int _{t}^{t+\Delta t}\rmd t\phi ^{2} (t')/2\lambda  }\times\nonumber\\
&&
{\mathcal T}\rme^{\rmi\int _{t}^{t+\Delta t}\phi (t')\hat{A}(t')\rmd t } \; \hat{\tilde{\rho }}\left(t\right)\;
\tilde{{\mathcal T}}\rme^{-\rmi\int _{t}^{t+\Delta t}\phi (t')\hat{A}(t')\rmd t } \rme^{(\rmi/\hbar)\hat{H}(t+\Delta t)}
 \end{eqnarray} 
Next use
\begin{eqnarray} \label{61)} 
&&{\mathcal T}\rme^{\rmi\int _{t}^{t+\Delta t}\phi (t')\hat{A}(t')\rmd t } ={\mathcal T}
\exp\left(\rmi\Delta t\phi(t)\rme^{\left(\rmi/\hbar \right)\hat{H}t } \hat{A}
 \rme^{-\left(\rmi/\hbar \right)\hat{H} }\right)  =\\
&&\prod _{j}\rme^{\left(\rmi/\hbar \right)\hat{H}t_{j} } \rme^{\rmi\Delta t\phi(t)\hat{A}} \rme^{-\left(\rmi/\hbar \right)\hat{H}t_{j} }
=\rme^{\left(\rmi/\hbar \right)\hat{H}\left(t+\Delta t\right)} \rme^{\int _{t}^{t+\Delta t}\left(\rmi/\hbar \right)\left[\phi (t')\hat{A}-\hat{H}\right]\rmd t' } \rme^{-\left(\rmi/\hbar \right)\hat{H}t}\nonumber
\end{eqnarray} 
to rewrite (\ref{rrr}) in the form
\begin{eqnarray} \label{rr1} 
&&\hat{\tilde{\rho }}\left(t+\Delta t\right)=\int  \rmD\phi {\kern 1pt} \rme^{-\int _{t}^{t+\Delta t}\rmd t\phi ^{2} (t)/2\lambda  }\times\nonumber\\
&&
\rme^{\rmi\int _{t}^{t+\Delta t}\left(\phi (t')\hat{A}-\hat{H}\right)\rmd t } \; \hat{\tilde{\rho }}\left(t\right)\; \rme^{-\rmi\int _{t}^{t+\Delta t}\left(\phi (t')\hat{A}-\hat{H}\right)\rmd t }.
\end{eqnarray} 
We next expand the RHS of (\ref{rr1}), keeping only terms that can contribute to order $O\left(\Delta t\right)$. To this end, we use
\begin{eqnarray} \label{63)} 
&&\rme^{\pm \rmi\int _{t}^{t+\Delta t} \rmd t\; \left(\phi (t)\hat{A}-\hat{H}/\hbar \right)} =1\pm \rmi\hat{A}\int _{t}^{t+\Delta t} \phi (t)\rmd t
\mp \rmi\left(\hat{H}/\hbar \right)\Delta t\nonumber\\
&&-\hat{A}^{2} \int _{t}^{t+\Delta t} \int _{t}^{t+\Delta t} \rmd t\rmd t'\phi (t)\phi (t')/2 
\end{eqnarray} 
This leads, using $\langle \phi \rangle =0$ and $\langle \phi (t)\phi (t')\rangle =\lambda \delta (t-t')$, to
\begin{eqnarray}
&&\hat{\tilde{\rho }}(t+\Delta t)=\hat{\tilde{\rho }}(t)
-\Delta t[\hat{H},\hat{\tilde{\rho }}(t)]\left(\rmi/\hbar \right)-\rmi\left[\hat{A},\hat{\tilde{\rho }}(t)\right]\int _{t}^{t+\Delta t} \langle \phi (t)\rangle \rmd t
 \nonumber\\
&&+\int _{t}^{t+\Delta t} \langle \phi (t)\phi (t')\rangle( \hat{A}\hat{\tilde{\rho }}(t)\hat{A}-
\{ \hat{A}^{2} ,\hat{\tilde{\rho }}(t)\} /2)
=\nonumber\\
&&\hat{\rho }(t)-\Delta t[\hat{H},\hat{\tilde{\rho }}(t)]\left(\rmi/\hbar \right)
-\lambda \Delta t\{ \hat{A}^{2} ,\hat{\tilde{\rho }}(t)\} /2+\lambda \Delta t\hat{A}\hat{\tilde{\rho }}(t)\hat{A},
\end{eqnarray}
which yields
\begin{eqnarray} \label{64)} 
&&\frac{\rmd\hat{\tilde{\rho }}(t)}{\rmd t} =-\frac{\rmi}{\hbar } [\hat{H},\hat{\tilde{\rho }}(t)]-\lambda \{ \hat{A}^{2} ,\hat{\tilde{\rho }}(t)\} /2+\lambda \hat{A}\hat{\tilde{\rho }}(t)\hat{A}\nonumber\\
&&=[\hat{H},\hat{\tilde{\rho }}(t)]/\rmi\hbar -\lambda [\hat{A},[\hat{A},\hat{\tilde{\rho }}(t)]]/2.
 \end{eqnarray}

\section*{Appendix F} 
\renewcommand{\theequation}{F.\arabic{equation}}
\setcounter{equation}{0}

Here we derive (\ref{avu1}-\ref{avu3}).
We will demonstrate the derivation of the two-time correlation function, (\ref{avu2}). We start from (\ref{av2})
 in the form (\ref{corr}) and use the cyclic permutation property of the trace together with identities such as
\begin{equation} \label{65)} 
{\mathcal T}\rme^{(\rmi/ \hbar)\int _{0}^{t'}\rmd\tau \hat{H}_{\phi } \left(\tau \right)}
\tilde{\mathcal T}\rme^{(\rmi/\hbar)\int _{0}^{t}\rmd\tau \hat{H}_{\phi } \left(\tau \right)}
= {\mathcal T}\rme^{(\rmi/\hbar)\int _{t}^{t'}\rmd\tau \hat{H}_{\phi } \left(\tau \right)}\; (t'>t) 
\end{equation} 
to get
\begin{eqnarray} \label{aac} 
&&\langle a(t)a(t')\rangle _{b} =\frac{1}{2} \int  \rmD_{\lambda } \phi\,
{\rm Tr}\left[\hat{A}\tilde{{\mathcal T}}\rme^{\int _{t}^{t'}\hat{H}_{\phi } (s)\rmd s/\rmi\hbar  }\times\right.\nonumber\\
&&\left.\left\{\hat{A},\tilde{{\mathcal T}}\rme^{\int _{0}^{t}\hat{H}_{\phi } (s)\rmd s/\rmi\hbar  } \hat{\rho }{\mathcal T}\rme^{\int _{0}^{t}\rmi\hat{H}_{\phi } (s)\rmd s/\hbar  } \right\}{\mathcal T}\rme^{\int _{t}^{t'}\rmi\hat{H}_{\phi } (s)\rmd s/\hbar  } \right],
\end{eqnarray} 
where $\hat{H}_{\phi } (t)=\hat{H}-\hbar \phi (t)\hat{A}$. The functional integral $\int \rmD_{\lambda } \phi  $ can be divided into a product of integrals performed over trajectories $\phi(t)$ between $0$ and $t$ and between $t$ and $t'$. The former operates only on the \textit{$\phi$} dependent expression in the anticommutator brackets, yielding $\int \rmD_{\lambda } \phi \, \tilde{{\mathcal T}}\rme^{-\int _{0}^{t}\rmi\hat{H}_{\phi } \left(s\right)\rmd s/\hbar  } \hat{\rho }{\mathcal T}\rme^{\int _{0}^{t}\rmi\hat{H}_{\phi } \left(s\right)\rmd s/\hbar  }  $ which satisfies
\begin{equation} \label{uuu} 
\int \rmD_{\lambda } \phi \, \tilde{{\mathcal T}}\rme^{-\int _{0}^{t}\rmi\hat{H}_{\phi } \left(s\right)\rmd s/\hbar  } \hat{\rho }{\mathcal T}\rme^{\int _{0}^{t}\rmi\hat{H}_{\phi } \left(s\right)\rmd s/\hbar  } 
=\breve{U}\left(t,0\right)\hat{\rho }=\hat{\tilde{\rho }}\left(t\right) 
\end{equation} 
((\ref{uuu}) is equivalent to (\ref{rr1}), generalized for finite time evolution). (\ref{aac}) becomes
\begin{equation} \label{68)} 
\langle a(t)a(t')\rangle _{b} =\int  \rmD_{\lambda } \phi {\rm Tr}\left[\hat{A}\tilde{{\mathcal T}}\rme^{-\rmi\int _{t}^{t'}\hat{H}_{\phi } (s)\rmd s/\hbar  }
\breve{A}\breve{U}\left(t,0\right)\hat{\rho }{\mathcal T}\rme^{\rmi\int _{t}^{t'}\hat{H}_{\phi } (s)\rmd s/\hbar  } \right] 
\end{equation} 
Using (\ref{uuu}) again, now in the form $\int \rmD_{\lambda } \phi \, \tilde{{\mathcal T}}\rme^{-\int _{t}^{t'}\rmi\hat{H}_{\phi } \left(s\right)\rmd s/\hbar  } \hat{x}\left(t\right){\mathcal T}\rme^{\int _{t}^{t'}\rmi\hat{H}_{\phi } \left(s\right)\rmd s/\hbar  }  =\breve{U}\left(t',t\right)\hat{x}\left(t\right)$, leads to (\ref{avu2}). (\ref{avu3}) is verified analogously.

\section*{Appendix G}
\renewcommand{\theequation}{G.\arabic{equation}}
\setcounter{equation}{0}
Here we prove (\ref{avs12}).
 From (\ref{avu1}), (\ref{sri}) and (\ref{suu}), using also $\breve{U}\hat{1}=1$ (since $\breve{L}\hat{1}=0$) and $\left\{\hat{\sigma }_{j} ,\hat{\sigma }_{k} \right\}=2\delta _{jk} \hat{1}$ we get
\begin{equation} \label{69)} 
\langle \sigma _{z} (t)\rangle _{b} =\frac{1}{2} {\rm Tr}\left[\hat{\sigma }_{z} \left(1+\hat{\sigma }_{z} \left(t\right)\right)\right]=\frac{z(t)}{2}{\rm Tr}\left(\hat{\sigma }_{z} ^{2} \right)=z\left(t\right).
\end{equation} 
 Next, from (\ref{avu2}) and (\ref{sri})
\begin{equation} \label{70)} 
\langle \sigma _{z} (t)\sigma _{z} (t')\rangle _{b} =\frac{1}{4}{\rm Tr}\left[\hat{\sigma }_{z} \breve{U}(t',t)\left\{\hat{\sigma }_{z} ,\left(\hat{1}+\hat{\sigma }_{z} \left(t\right)\right)\right\}\right]
\end{equation} 
for $t'>t$.
Using (\ref{suu}) gives $\left\{\hat{\sigma }_{z} ,\left(\hat{1}+\sigma _{z} \left(t\right)\right)/2\right\}=\hat{\sigma }_{z} +z\left(t\right)\hat{1}$. Finally,
\begin{eqnarray} \label{71)} 
&&\langle \sigma _{z} (t)\sigma _{z} (t')\rangle _{b} =\frac{1}{2} {\rm Tr}\left[\hat{\sigma }_{z} \breve{U}(t',t)\left(\hat{\sigma }_{z} +z\left(t\right)\hat{1}\right)\right]\nonumber\\
&&
=\frac{1}{2} {\rm Tr}\left[\hat{\sigma }_{z} \left(\hat{\sigma }_{z} \left(t'-t\right)+z\left(t\right)\hat{1}\right)\right]=z\left(t'-t\right) 
\end{eqnarray}

\end{document}